\documentclass[preprint2]{aastex}
\usepackage{graphicx}
\usepackage{graphics}
\usepackage{color}
\DeclareGraphicsExtensions{.pdf}

\slugcomment{Accepted for publication in the Astrophysical Journal}

\shortauthors{Halverson et al.}
\shorttitle{Bullet Cluster SZE Observations with APEX-SZ}

\begin{document}

\newcommand{\oscdemod}{oscillator/demodulator}
\newcommand{\fMUX}{fMUX}
\newcommand{\squid}{SQUID}
\newcommand{\rtHz}{$\sqrt{\mbox{Hz}}$}
\newcommand{\phinot}{\mbox{$\Phi_0$}}
\newcommand{\degree}{\mbox{$^{\circ}$}}
\newcommand{\fortran}{{\tt Fortran~77}}
\newcommand{\CXX}{C++}
\newcommand{\order}{\mbox{${\cal O}$}}
\newcommand{\const}{\mbox{\sc\small Const}}
\newcommand{\mycomment}[1]{{\bf\it\color{red} #1}} 


\title{Sunyaev-Zel'dovich Effect Observations of the Bullet Cluster (1E 0657--56) with APEX-SZ}


\author{N.~W.~Halverson,\altaffilmark{1,2} T.~Lanting,\altaffilmark{3}
P.~A.~R.~Ade,\altaffilmark{4} K.~Basu,\altaffilmark{5}
A.~N.~Bender,\altaffilmark{1}
B.~A.~Benson,\altaffilmark{6}
F.~Bertoldi,\altaffilmark{5}
H.-M.~Cho,\altaffilmark{7} G.~Chon,\altaffilmark{8}
J.~Clarke,\altaffilmark{6,9} M.~Dobbs,\altaffilmark{3}
D.~Ferrusca,\altaffilmark{6} R.~G\"usten,\altaffilmark{8}
W.~L.~Holzapfel,\altaffilmark{6} A.
Kov\'acs,\altaffilmark{8} J.~Kennedy,\altaffilmark{3}
Z.~Kermish,\altaffilmark{6} R.~Kneissl,\altaffilmark{8}
A.~T.~Lee,\altaffilmark{6,9} M.~Lueker,\altaffilmark{6}
J.~Mehl,\altaffilmark{6} K.~M.~Menten,\altaffilmark{8}
D.~Muders,\altaffilmark{8} M.~Nord,\altaffilmark{5,8}
F.~Pacaud,\altaffilmark{5} T.~Plagge,\altaffilmark{6}
C.~Reichardt,\altaffilmark{6} P.~L.~Richards,\altaffilmark{6}
R.~Schaaf,\altaffilmark{5} P.~Schilke,\altaffilmark{8}
F.~Schuller,\altaffilmark{8} D.~Schwan,\altaffilmark{6}
H.~Spieler,\altaffilmark{9} C.~Tucker,\altaffilmark{4}
A.~Weiss,\altaffilmark{8} O.~Zahn\altaffilmark{6}}

\altaffiltext{1} {Center for Astrophysics and Space Astronomy, Department of Astrophysical and Planetary Sciences, University of Colorado, Boulder, CO, 80309}
\altaffiltext{2} {Department of Physics, University of Colorado, Boulder, CO, 80309}
\altaffiltext{3}{Department of Physics, McGill University, Montr\'{e}al, Canada, H3A 2T8}
\altaffiltext{4}{School of Physics and Astronomy, Cardiff University, CF24 3YB Wales, UK}
\altaffiltext{5}{Argelander Institute for Astronomy, Bonn University, Bonn, Germany}
\altaffiltext{6}{Department of Physics, University of California, Berkeley, CA, 94720}
\altaffiltext{7}{National Institute of Standards and Technology, Boulder, CO, 80305}
\altaffiltext{8}{Max Planck Institute for Radioastronomy, 53121 Bonn, Germany}
\altaffiltext{9}{Lawrence Berkeley National Laboratory, Berkeley, CA, 94720}

\begin{abstract}
We present observations of  the Sunyaev-Zel'dovich effect (SZE) in
the Bullet cluster (1E 0657--56) using the APEX-SZ instrument at 150$\,$GHz
with a resolution of $1\arcmin$. The main results are maps of the
SZE in this massive, merging galaxy cluster. The cluster is
detected with $23\,\sigma$ significance
within the central 1\arcmin\ radius of the source position.
The SZE map has a broadly similar morphology to that in existing X-ray
maps of this system, and
we find no evidence for significant contamination of the SZE emission by
radio or IR sources.
In order to make simple quantitative comparisons with cluster
gas models derived from X-ray observations,
we fit our data to an isothermal elliptical $\beta$ model, despite the
inadequacy of such a model for this complex merging system.
With an X-ray derived prior on the power-law index, $\beta = 1.04^{+0.16}_{-0.10}$,
we find a core radius $r_c =142 \pm 18 \arcsec$, an axial
ratio of $0.889 \pm 0.072$, and a central temperature decrement
of $-771 \pm 71\,\mu{\rm K_{CMB}}$, including a $\pm 5.5\%$ flux calibration uncertainty.
Combining the APEX-SZ map with a map of projected electron surface density from Chandra X-ray
observations, we determine the mass-weighted temperature of the cluster gas to be
$T_{mg}=10.8 \pm 0.9\,$keV, significantly lower than some previously reported X-ray spectroscopic
temperatures.
Under the assumption of an isothermal cluster gas distribution in hydrostatic equilibrium,
we compute the gas mass fraction for prolate and oblate spheroidal geometries and find it to
be consistent with previous results from X-ray and weak lensing observations.
This work is the first result from the APEX-SZ experiment, and
represents the first reported scientific result from
observations with a large array of multiplexed superconducting
transition-edge sensor bolometers.
\end{abstract}

\keywords{cosmic microwave background --- cosmology:observations ---
galaxies: clusters: individual (1E 0657--56)}



\section{Introduction}
\label{SEC:introduction}

Clusters of galaxies are a unique probe of the growth and dynamics of
structure in the Universe.
In particular, active mergers of sub-clusters provide a window to the
processes by which massive clusters are assembled.
In these systems, the galaxies and associated dark
matter are essentially collisionless. In contrast, the ionized
intracluster gas, typically at temperatures of $T \sim 10^8$~K,
is strongly interacting and experiences complex dynamics.
In extreme cases, the normally associated dark matter and
intracluster gas can be significantly separated.

The Bullet cluster (1E 0657--56) at $z=0.296$, is a massive cluster
consisting of two
sub-clusters in the process of merging. The smaller sub-cluster or
``bullet" has passed through the larger main cluster.
X-ray observations infer a bow shock velocity of $\sim$4700 km/s~\markcite{markevitch2006}({Markevitch} 2006),
while simulations of the collision yield a substantially lower speed for the
sub-cluster \markcite{springel2007}({Springel} \& {Farrar} 2007).
This collision is perpendicular to the line of sight,
providing an ideal system for studying interacting sub-clusters~\markcite{clowe2006}({Clowe} {et~al.} 2006).

The mass surface density of the Bullet cluster has been measured using
weak and strong gravitational lensing of light from background
galaxies. There are significant angular offsets between the peaks of
the X-ray surface brightness, which trace the baryonic gas through
thermal bremsstrahlung emission,
and the peaks of the lensing surface
density, which are associated with the majority of the mass. The
combined weak and strong lensing analyses of \markcite{bradac2006}{Brada{\v c}} {et~al.} (2006) show
that the main cluster and sub-cluster are separated from their
associated X-ray peaks at $10\,\sigma$ and $6\,\sigma$ significance
respectively. This result has been recognized as providing direct
evidence for the presence of collisionless dark matter in this
system~\markcite{clowe2006}({Clowe} {et~al.} 2006).

The Sunyaev-Zel'dovich effect (SZE) provides an independent probe of the intracluster gas.
In the SZE, a small fraction
($\sim1\%$) of cosmic microwave background (CMB) photons undergo
inverse Compton scattering from intracluster
electrons~\markcite{sunyaev1970,birkinshaw1999}({Sunyaev} \& {Zel'dovich} 1970; {Birkinshaw} 1999). This process distorts
the Planck blackbody spectrum of the CMB and produces a signal
proportional to the gas pressure integrated along the line of sight.
At $150\,$GHz, the SZE produces a temperature decrement with respect to the
unperturbed CMB intensity.
Early detections of the SZE in the Bullet cluster include
work by \markcite{andreani1999}{Andreani} {et~al.} (1999) and \markcite{gomez2003}{Gomez} {et~al.} (2004).

Unlike the X-ray surface brightness, the peak SZE surface brightness for a
given cluster is independent of redshift.
Therefore, the SZE has the potential to be an effective probe of intracluster gas
out to the redshifts at which clusters are assembled.
SZE measurements of galaxy clusters provide
complementary constraints on cluster properties typically derived
from X-ray measurements such as central electron density, core
radius of the intracluster gas, cluster gas mass, and fraction of the total cluster mass
in gas. Since the SZE and X-ray signals are proportional to the line-of-sight integral of
the electron density and electron density squared respectively, SZE results will be less
sensitive to clumping of the intracluster gas.
For all comparisons between SZ and X-ray data, we assume
a $\Lambda$CDM cosmology, with $h = 0.7$, $\Omega_{\rm{m}} = 0.27$, and $\Omega_{\Lambda} = 0.73$.

In this paper, we present a 1\arcmin\ resolution SZE image of the
Bullet cluster at $150\,$GHz made with the APEX-SZ instrument. It is the
first reported scientific result from observations with a large array of
multiplexed superconducting transition-edge sensor bolometers. In
\S\,\ref{SEC:observations}, we discuss the instrument and
observations. Calibration is discussed in \S\,\ref{SEC:calibration}.
In \S\,\ref{SEC:datareduction}, we describe the data
reduction procedure, and in \S\,\ref{SEC:results}, we present the
results of fits to the SZE surface brightness with cluster models,
including mass-weighted electron temperature and gas mass 
fraction calculations.
We summarize the conclusions and discuss
future work in \S\,\ref{SEC:conclusions}.

\section{Observations}
\label{SEC:observations}

APEX-SZ is a receiver designed specifically for SZE galaxy cluster
surveys~\markcite{schwan2003,dobbs2006,schwan2008}({Schwan} {et~al.} 2003; {Dobbs} {et~al.} 2006; Schwan {et~al.} 2009, in preparation). It
is mounted on the 12-meter diameter APEX telescope, located on the
Atacama plateau in northern Chile~\markcite{guesten2006}({G{\"u}sten} {et~al.} 2006). The observing
site was chosen for its extremely dry and stable atmospheric
conditions. The mean atmospheric transmittance is frequently better
than 95$\%$ in the APEX-SZ frequency band at
150~GHz~\markcite{peterson2003,chamberlain1995}({Peterson} {et~al.} 2003; {Chamberlain} \& {Bally} 1995). The telescope is capable
of round-the-clock observations.

Three reimaging mirrors in the Cassegrain cabin couple the APEX
telescope to the focal plane of APEX-SZ. We achieve the diffraction
limited performance of the telescope across the entire 0.4\degree\
field of view with a mean measured beam full width half maximum (FWHM)
of 58\arcsec, and a measured beam solid angle of 1.5~arcmin$^2$,
including measured sidelobes at the $-14$~dB level.

The APEX-SZ receiver houses a cryogenic focal plane, operating at 0.3~K.
The focal plane contains 330 horn-fed absorber-coupled superconducting
transition-edge sensor
bolometers~\markcite{ref:richards-bolometer-review,lee1996}(Richards 1994; {Lee} {et~al.} 1996), 
with 55 detectors on each of six sub-array wafers. Of the 330
detectors, 280 are read out with the current frequency-domain multiplexed
readout hardware. We measure
the median individual pixel Noise Equivalent Power (NEP) to be $10^{-16}
$~W/$\sqrt{{\rm Hz}}$ and the median Noise Equivalent Temperature (NET)
to be $860~\mu$K$_{\rm{CMB}}\sqrt{{\rm s}}$.
The measured optical bandwidth of the receiver is 40\% narrower
than the design goal of 38~GHz, resulting in lower sensitivity than
anticipated.

\begin{figure}[h]\centering
\includegraphics[width=0.45\textwidth]{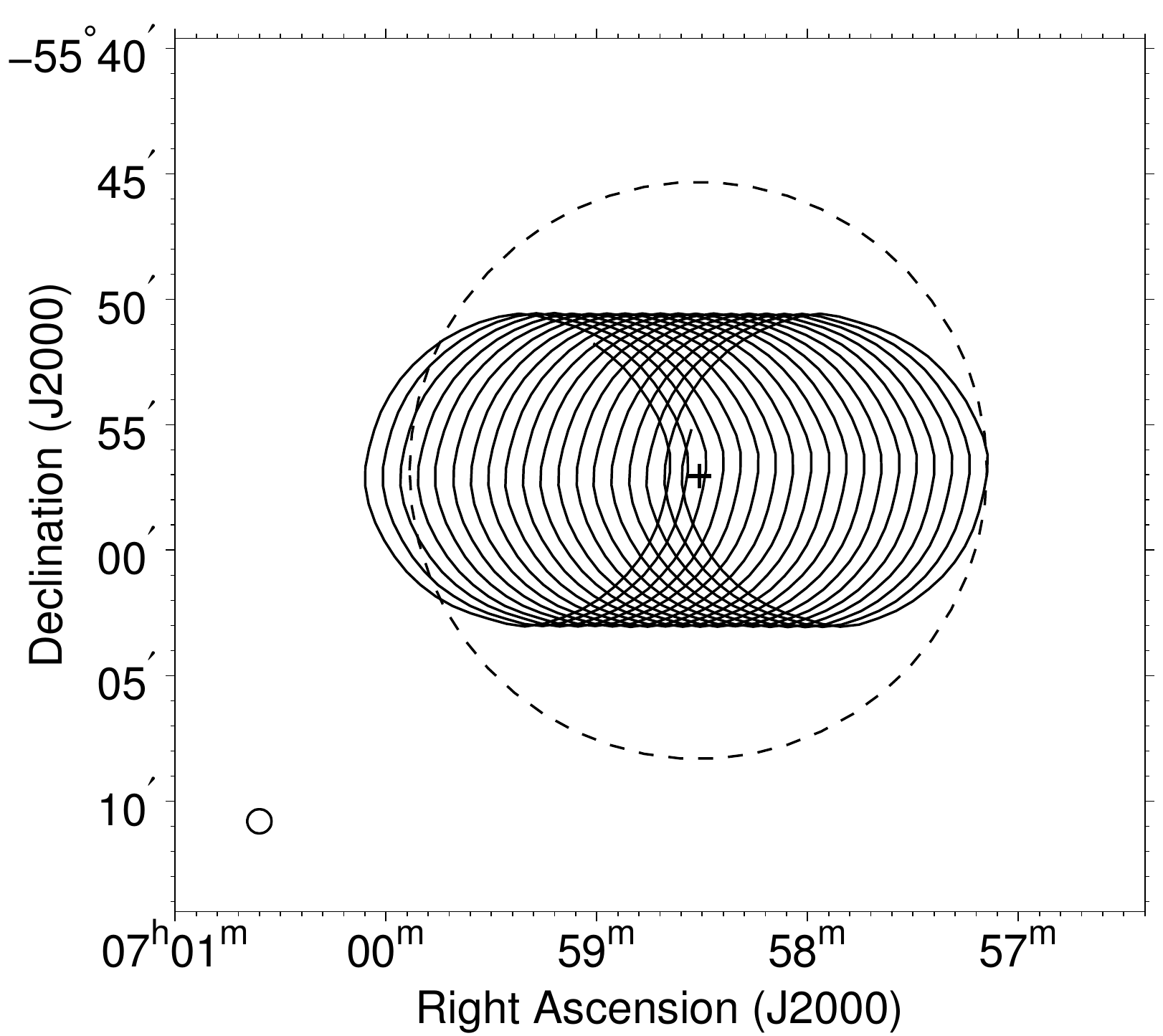}
  \caption[]{ The 100-second circular drift scan pattern. The solid
   line shows the track of the center of the array. One circle has a
   period of five seconds. The dashed line shows the instantaneous
   field of view of the bolometer array. The $+$ marker indicates the source
   position with respect to the scan pattern. The center of the
   circles is constant in azimuth and elevation as the source drifts
   across the field. The small disk in the lower left indicates the
   58\arcsec\ mean FWHM beam for a single bolometer.}
\label{FIG:scanpattern}
\end{figure}

The large field of view of the APEX-SZ instrument is designed for
surveying large areas of sky. In order to efficiently observe a single
target, we use the circular scan pattern illustrated in
Figure~\ref{FIG:scanpattern}. The circle center is fixed in AZ/EL
coordinates for twenty circular sub-scans, with a total duration of 100
seconds. This choice has a number of important advantages.
The sky signal is modulated so that it appears in the timestream at frequencies
higher than atmospheric drifts and readout $1/f$ noise.
In addition, the circle scan has a moderate continuous acceleration; the lack
of high acceleration turn-arounds makes it possible to achieve a high observing
efficiency.
Approximately $20\%$ of the total observing time is spent moving the telescope
to a new center position before the start of the next scan.
Every bolometer maps a $12\arcmin \times 25\arcmin$ sub-field,
with a combined map field of $36\arcmin \times 48\arcmin$ every 100
seconds.

Observations of the Bullet cluster were conducted over a period of
seven days in August 2007, when the cluster was visible between the
hours of 03:00 and 15:00 local time. The weather over this period was
typical for the site, with precipitable water vapor varying between
0.25 and 1.5 mm, and a median atmospheric transmittance of $97\%$. For
the analysis in this paper, 235 scans are used, each scan consisting
of twenty 5-s circular sub-scans, for a total of 6.4 hours of on-source
data.

\section{Calibration}
\label{SEC:calibration}

The response of the receiver to astronomical sources is measured with
daily raster scans of Mars over every bolometer in the
array. For each bolometer, the observations provide a primary flux
calibration and a high signal-to-noise beam profile from which we
determine beam parameters such as size, ellipticity, and position with
respect to the array-center pointing.
Additional observations of RCW57
and RCW38 are used to monitor gain stability, and frequent
observations of bright quasars near the cluster source are used to
monitor pointing stability.

The WMAP satellite has been used to calibrate the brightness
temperature of Mars at $93\,$GHz in five measurement periods spanning
several years \markcite{hill08}({Hill} {et~al.} 2009). The WMAP Mars temperatures are tied to
the CMB dipole moment and are accurate to better than $1.0\%$.  The
brightness temperature of Mars changes significantly ($\sim 15\%$) as
a function of its orbit and orientation.  We use a version of the Rudy
Model \markcite{rudy87,muhleman91}({Rudy} {et~al.} 1987; {Muhleman} \& {Berge} 1991), that has been updated and maintained
by Bryan
Butler,\footnote{http://www.aoc.nrao.edu/\~{}bbutler/work/mars/model/}
to transfer the WMAP Mars temperature results to the APEX-SZ frequency
band and specific times of our Mars observations.

After applying a constant scaling factor, we find the Rudy model predictions
for the Mars brightness temperature to be in excellent agreement with
the WMAP measurements.
We find that the Rudy model
brightness temperatures at $93\,$GHz are systematically a factor of
$1.052 \pm 0.010$ higher than those measured by WMAP in the five
published observation periods.
In contrast, repeating the same exercise with the thermal model
developed by \markcite{wright76,wright07}{Wright} (1976, 2007), as implemented in the
online JCMT-FLUXES program,\footnote{http://www.jach.hawaii.edu/jac-bin/planetflux.pl}
results in a scaling factor of $1.085\pm 0.043$.
This is consistent with the 10\% rescaling of this model
called for in \markcite{hill08}{Hill} {et~al.} (2009), but the scaling factor exhibits significantly
larger rms scatter than that of the Rudy Model.

We therefore use the WMAP $93\,$GHz calibrated Rudy Model to compute
the Mars brightness temperatures at $150\,$GHz for the specific times
of our Mars observations by reducing the Rudy model $150\,$GHz
temperatures by a factor of 1.052. The Rudy model $93\,$GHz to $150\,$GHz frequency scaling factor
is $1.016 \pm 0.009$ at the times of our Mars observations, and we
adopt the rms scatter in this frequency scaling factor as an estimate of its
uncertainty. Combining the uncertainties in the WMAP Mars calibration,
the WMAP to Rudy model scaling factor at $93\,$GHz, and the Rudy model frequency
scaling factor, we estimate the uncertainty in Mars temperature to be $\pm 1.7\%$.

The measured signals from the calibrators are corrected for atmospheric
opacity, which is measured with a sky dip observation at the beginning and end of each day's
observations. Measured zenith transmittance over the observing period ranged
between 0.92 and 0.98, with a median of 0.97. Based on the
observed temporal variability of
the opacity, drifts in atmospheric opacity between the sky dip and
observation contribute $<0.4\%$ to the overall calibration uncertainty.
After correcting for the atmospheric opacity, we find that the
Mars temperature measured by APEX-SZ varies from the model prediction
by up to $\sim 3\%$ over the course of the observation period.
This gain variation is included as a source of error
in the final calibration uncertainty.
The APEX-SZ observing band center is measured with a Fourier
transform spectrometer to be $152\pm2$ GHz.
The uncertainty in the band center results in a $\pm 1.4\%$ uncertainty
in extrapolation of the Mars based calibration to CMB temperature.

The beam shape, including near sidelobes, is characterized by creating
a beam map from Mars observations, combining the same bolometer
channels that are used to make the science maps. We adjust the
calibration and measured beam size for the small ($\sim 1\%$)
correction due to the $8^{\prime \prime}$ angular size of Mars. We estimate
a fractional uncertainty in the beam solid angle of $\pm 4\%$.

The APEX-SZ detectors operate in a state of strong negative
electrothermal feedback which results in a linear response to changes
in the input optical power.
We have measured the response of the detectors during sky dips
between $90^\circ$ and $30^\circ$ elevation (antenna temperature
difference $\sim13\,$K),
and find no significant deviation from the expected linear response to loading.
We therefore conclude that detector
non-linearity makes a negligible contribution to the calibration
uncertainty.

Slowly changing errors in telescope pointing result in both a
pointing uncertainty and a flux calibration
uncertainty due a broadening of the effective beam pattern.
To measure pointing errors during our
observations, we observe a bright quasar within a few degrees of the
Bullet cluster every 1--2 hours, and apply a pointing correction
as needed.
The typical RMS pointing variations of the APEX telescope between
quasar observations is $\sim 4 \arcsec$.
This pointing uncertainty results in a slightly larger effective
beam for the coadded maps than is measured with the individual calibrator
maps.
The correction to the flux calibration of the coadded maps due to
pointing uncertainty is negligible,
particularly for the observation of extended objects such as the
Bullet cluster. We estimate the pointing uncertainty
in the coadded maps to be $\pm 4\arcsec$ in both
the RA and DEC directions.

The uncertainty in the CMB temperature calibration of the APEX-SZ
maps is summarized in Table~\ref{TBL:cal}.
The combination of all contributions to the calibration uncertainty
described above results in an overall point source flux uncertainty of
$\pm 5.5\%$.

\begin{deluxetable}{ll}
\tablecaption{\label{TBL:cal} APEX-SZ Flux Calibration Uncertainty}
\tablewidth{0pt}
\tablehead{
\colhead{Source} & \colhead{Uncertainty}}
\startdata
WMAP Mars temperature at $93\,$GHz & $\pm1.0\%$ \\
Rudy model to WMAP scaling factor at $93\,$GHz & $\pm1.0\%$ \\
$93\,$GHz to $150\,$GHz frequency scaling factor & $\pm0.9\%$ \\
Frequency band center & $\pm1.4\%$ \\
Beam solid angle & $\pm4.0\%$ \\
Atmospheric attenuation & $\pm0.4\%$ \\
Gain variation & $\pm3.0\%$ \\\hline
Total & $\pm5.5\%$ \\

\enddata
\end{deluxetable}

\section{Data Reduction}
\label{SEC:datareduction}

The data consist of 280 bolometer timestreams sampled at 100
Hz, telescope pointing data interpolated to the same rate,
housekeeping thermometry data, bolometer bias and readout
configuration data, and other miscellaneous monitoring data. The
fundamental observation unit is a scan comprising twenty 5-s
circular sub-scans in AZ/EL coordinates, allowing the source to drift
through the field of view (FOV), as described in \S\,\ref{SEC:observations} above.

Data reduction consists of cuts to remove
poor-quality data, filtering of $1/f$ and correlated noise due to
atmospheric fluctuations, and binning the bolometer data into maps.
These steps are described in more detail below.

\subsection{Timestream Data Cuts}

Timestream data are first parsed into individual circular sub-scans.
We reject $\sim 7$\% of the data at the beginning and end of the scan
where the telescope deviates from the constant angular velocity
circular pattern.
We reject bolometer channels that are optically or electronically unresponsive, or lack high-quality flux calibration data;
typically, 160--200 of the 280 bolometer channels remain after these
preliminary cuts. The large number of rejected channels is due primarily
to low fabrication yield for two of the six bolometer sub-array wafers.

We reject spikes and step-like glitches caused by cosmic rays or
electrical interference. These are infrequent and occur on time scales
faster than the detector optical time constant. We use a simple
signal-to-noise cut on the data to reject these, since the
timestream is noise dominated even for the $\sim$1~mK
Bullet cluster signal. Step-like glitches are often correlated
across many channels in the array, so we reject data from all channels whenever a
spike or glitch in $\geq 2\%$ of the channels is detected. These cuts result
in a loss of 8\% of the remaining data.

For each circular sub-scan, we also reject channels that have excessive
noise in signal band, resulting in a loss of 19\% of the remaining data.

\subsection{Atmospheric Fluctuation Removal}
\label{SEC:atmremoval}

After the timestream data cuts, fluctuations in atmospheric emission
produce the dominant signal in the raw bolometer timestreams. The
atmospheric signal is highly correlated across the array, which can be
exploited to remove the signal.
Principal component analysis (PCA) has been used by some groups
to reduce the atmospheric signal \markcite{scott2008,laurent2005}(see, e.g., {Scott} {et~al.} 2008; {Laurent} {et~al.} 2005).
However, the effect of PCA filtering on the source is difficult to predict and
is a function of the atmospheric conditions.
We have developed an analysis strategy that reduces the atmospheric signal
through the application of spatial filters that have a constant and well
understood effect on the signals we are attempting to measure.

Atmospheric fluctuation power is expected to follow a Kolmogorov
spatial power spectrum, with most power present on scales larger
than the separation between beams as they pass through the atmosphere,
resulting in an atmospheric signal that is highly correlated across the array.
To reduce these fluctuations, we first remove a polynomial and an elevation dependent
airmass opacity model from each channel's timestream, then remove
a first-order two-dimensional spatial polynomial across the
array for each time-step. This algorithm is described in detail
in the two following subsections. This atmospheric fluctuation removal
strategy requires that both the spatial extent of the scan
pattern and the instantaneous array FOV are larger than the
source. The $6\arcmin$ radius circular scan and the 23$\arcmin$ array
FOV allow us to recover most of the Bullet cluster's flux,
but some extended emission is lost as is described in \S\,\ref{SEC:results}.

\subsubsection{Timestream Atmosphere Removal}\label{subsec:circle}

We observe scan-synchronous signals in the bolometer timestreams due
to elevation-dependent atmospheric emission.  The optical path length
$L$ through the atmosphere is proportional to the cosecant of the
elevation angle $\epsilon$, $L \propto \csc(\epsilon)$. The change in
optical path-length is nearly a linear function of elevation angle
over the $6\arcmin$ circular scan radius. In the circular drift scans,
this modulation of the elevation-dependent opacity produces an
approximately sinusoidal modulation in the bolometer timestream. For
each channel in each scan, we simultaneously fit and remove an
atmospheric model consisting of this cosecant function plus an order
20 polynomial (one degree of freedom per circular sub-scan) to remove
slow drifts in the atmospheric opacity and readout $1/f$ noise.
This scheme effectively
removes the common scan-induced atmospheric signal as well as most of
the atmospheric fluctuation power below the frequency of the circular
sub-scan (0.20 Hz), while only modestly affecting the central Bullet
cluster signal.

\subsubsection{Spatially Correlated Atmosphere Removal}

Removing the cosecant-plus-polynomial model
from the timestream data reduces low-frequency
atmospheric fluctuation power, but not higher frequency power
corresponding to smaller spatial scales near those where the cluster
signal occurs. To reduce these fluctuations, the atmosphere can be modeled
as a spatially correlated signal across the array pixel positions on
the sky with a low-order two-dimensional polynomial function. At each
time step, we fit and remove a low-order
two-dimensional polynomial
function across the array, similar to the procedure described in
\markcite{sayers07}Sayers (2007).
The relative gain coefficients for each bolometer channel are
calculated by taking the ratio of each channel's timestream, which is
dominated by correlated atmospheric noise, to a median timestream
signal generated from all channels. With the favorable atmospheric
conditions of these observations, we find that a first order spatial
polynomial (offset and tilt) is adequate to remove most of the atmospheric signal while
preserving the cluster signal.

Bolometer channels with excess uncorrelated noise are
more easily identified after removing the correlated atmospheric
noise component; we reject these noisy channels, then perform the
spatially correlated signal removal a second time.

\subsection{Map-Making}
\label{SEC:mapmaking}

The atmospheric removal algorithms described above act as a high-pass
filter. They suppress signals on scales comparable to the scan length
or the focal plane FOV. The cluster emission can be quite
extended, and therefore the data reduction filtering process attenuates
diffuse flux in the cluster signal and produces small
positive sidelobes around the cluster decrement. The data reduction
pipeline filters can be tailored, within limits, to meet various
scientific objectives. Thus, our primary data products consist of two
different high signal-to-noise maps of the cluster.

For one map, we mask a circular region centered on the cluster source
prior to fitting the timestream and spatial filters described in
\S\,\ref{SEC:atmremoval}, then apply the resulting filter functions to
the entire data set, including the source region. The source-mask
procedure prevents the cluster signal within the masked region from
influencing the baseline fits, and thus reduces attenuation of the
source central decrement and extended emission at the expense of
increasing the contribution of low-frequency noise in the map
center. We choose a source-mask radius of $4.75\arcmin$ as a
compromise between attenuation of diffuse emission and increased map
noise. We use the source-masked map to visually interpret the
morphology and extended emission in the cluster. These results are discussed in \S\,\ref{subsec:tempmap}.

We also produce a map in which we do not mask the source when applying
filters. The non-source-masked
map is used for model parameter estimation because it has higher signal to noise
in the central region of the map. In addition, it is easier to take
into account the effects of the data reduction filters, or transfer
function, on the underlying sky intensity distribution, which is
necessary for comparing the data to the model for parameter
estimation. The fitting procedure and results are described in more detail in
\S\,\ref{subsec:fitting}.

For each of the two maps, the post-cut, filtered timestream data are
binned in angular sky coordinates to create maps. For a given scan, a
map is created from each bolometer channel, applying the channel's
pointing offset and flux calibration.
A coadded scan map is created by combining individual channel maps
with minimum variance weighting in each pixel, using the sample variance
of the conditioned timestream data in the scan.
The final coadded map
is created by combining all scan maps, again with minimum variance
weighting in each pixel. We bin maps at a resolution of 10\arcsec\ to
oversample the beam. The source-masked map that we present
in \S\,\ref{SEC:results} is
convolved with a 1\arcmin\ FWHM Gaussian to smooth noise fluctuations
to the angular size of the beam. However, the radial profiles
presented below and the non-source-masked map used for model
fitting do not include this additional smoothing.

\section{Results}
\label{SEC:results}

\begin{figure*}[p!]\centering
\includegraphics[width=0.7\textwidth]{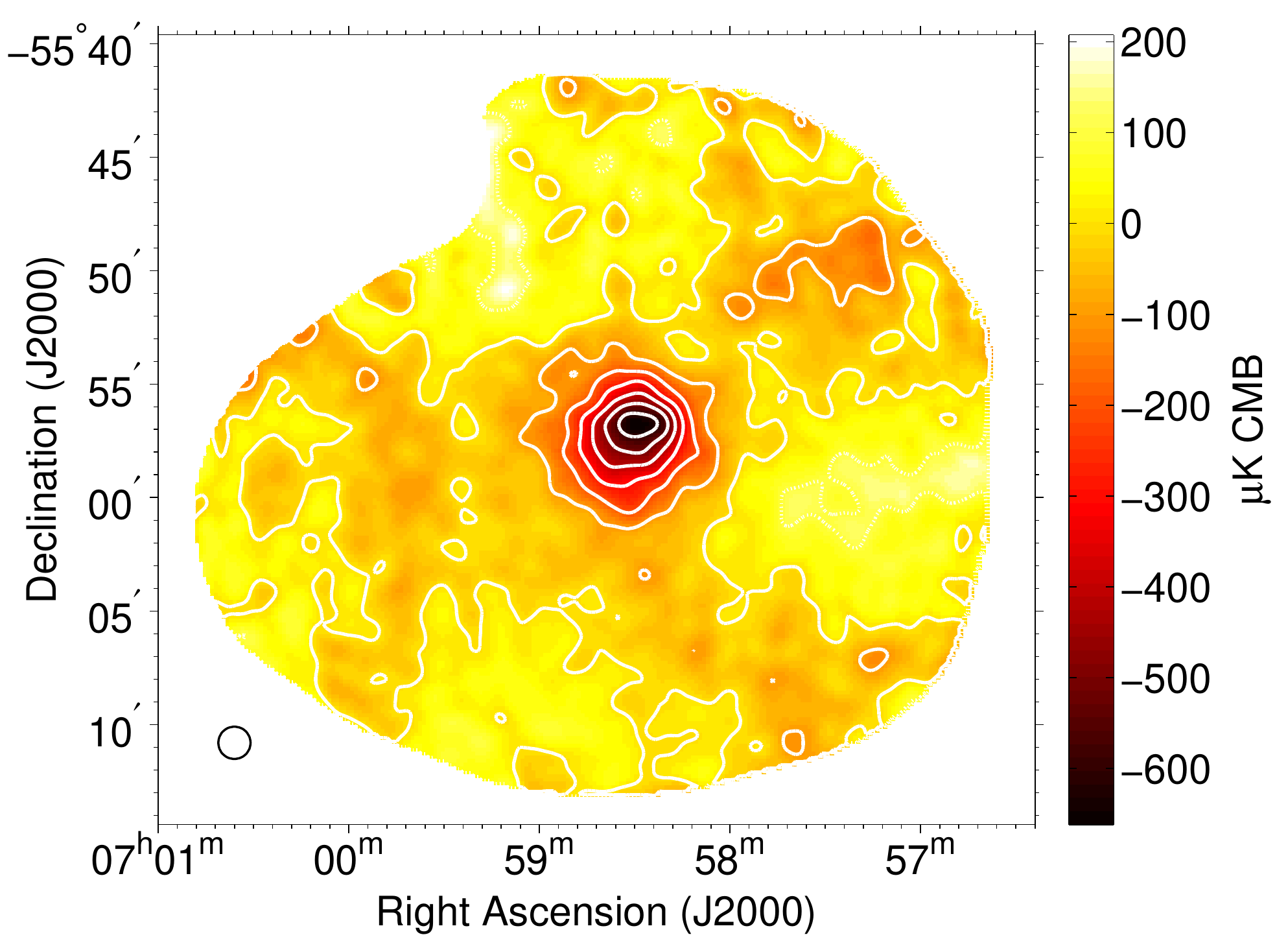}
\includegraphics[width=0.7\textwidth]{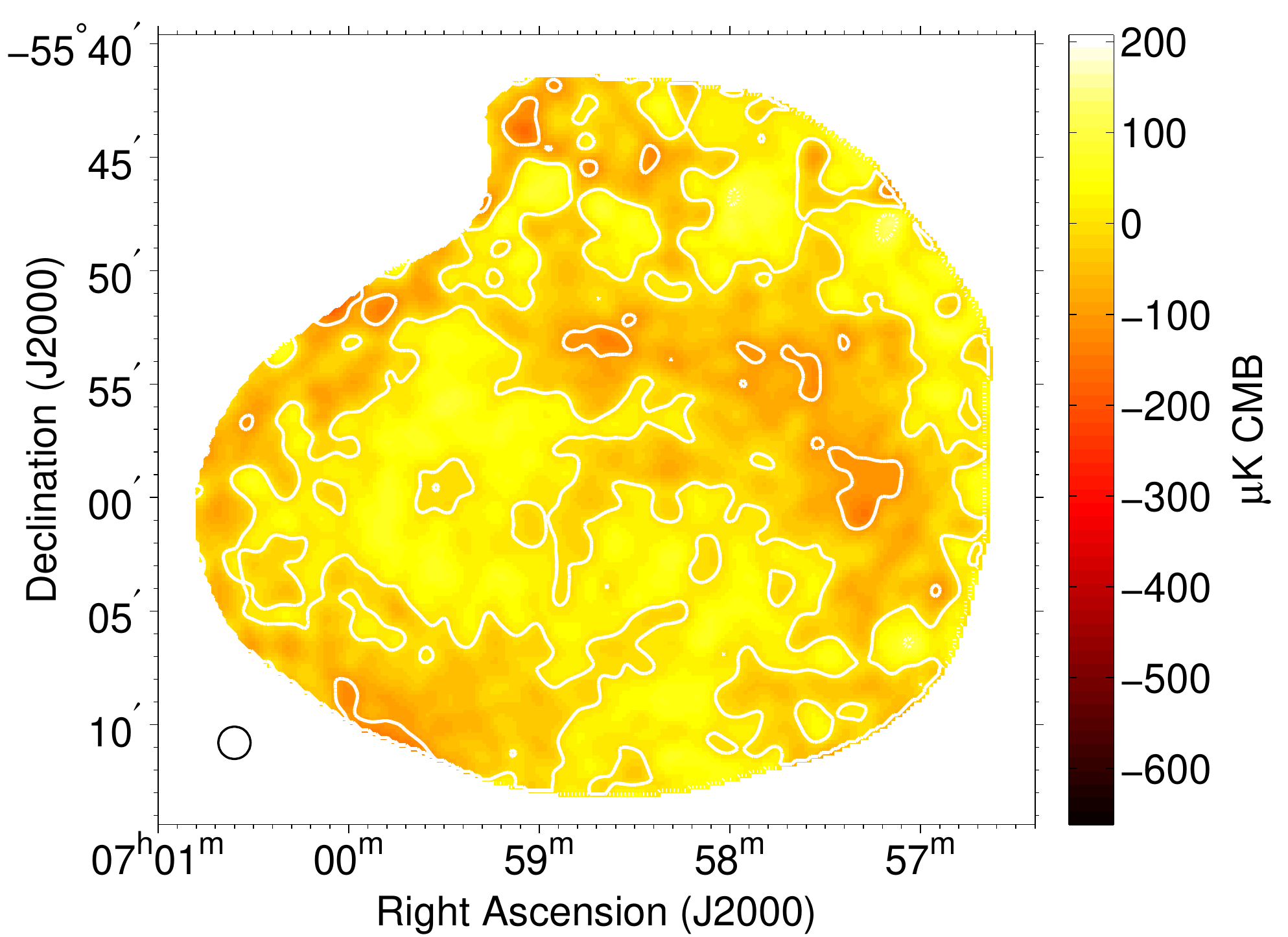}
  \caption[]{ (Top) Temperature map of the Bullet cluster system from
the source-masked data reduction, with scale in CMB temperature
units. The circle in the lower left corner represents the 85\arcsec\
FWHM map resolution which is the result of the instrument beam and
data reduction filter convolved with the 1\arcmin\ FWHM Gaussian
smoothing applied to the map. (Bottom) Difference map made by
multiplying alternate scan maps by $+1$ and $-1$,
respectively, then coadding all scan maps with minimum variance
weighting, in the same manner as was used to produce the temperature
map shown in the top panel. The contour interval is 100~$\mu \rm{K}_{\rm{CMB}}$
in both maps.}
  \label{FIG:bulletcoadd}
\end{figure*}

\subsection{Temperature Map}
\label{subsec:tempmap}

\begin{figure}[h!]\centering
\includegraphics[width=0.45\textwidth]{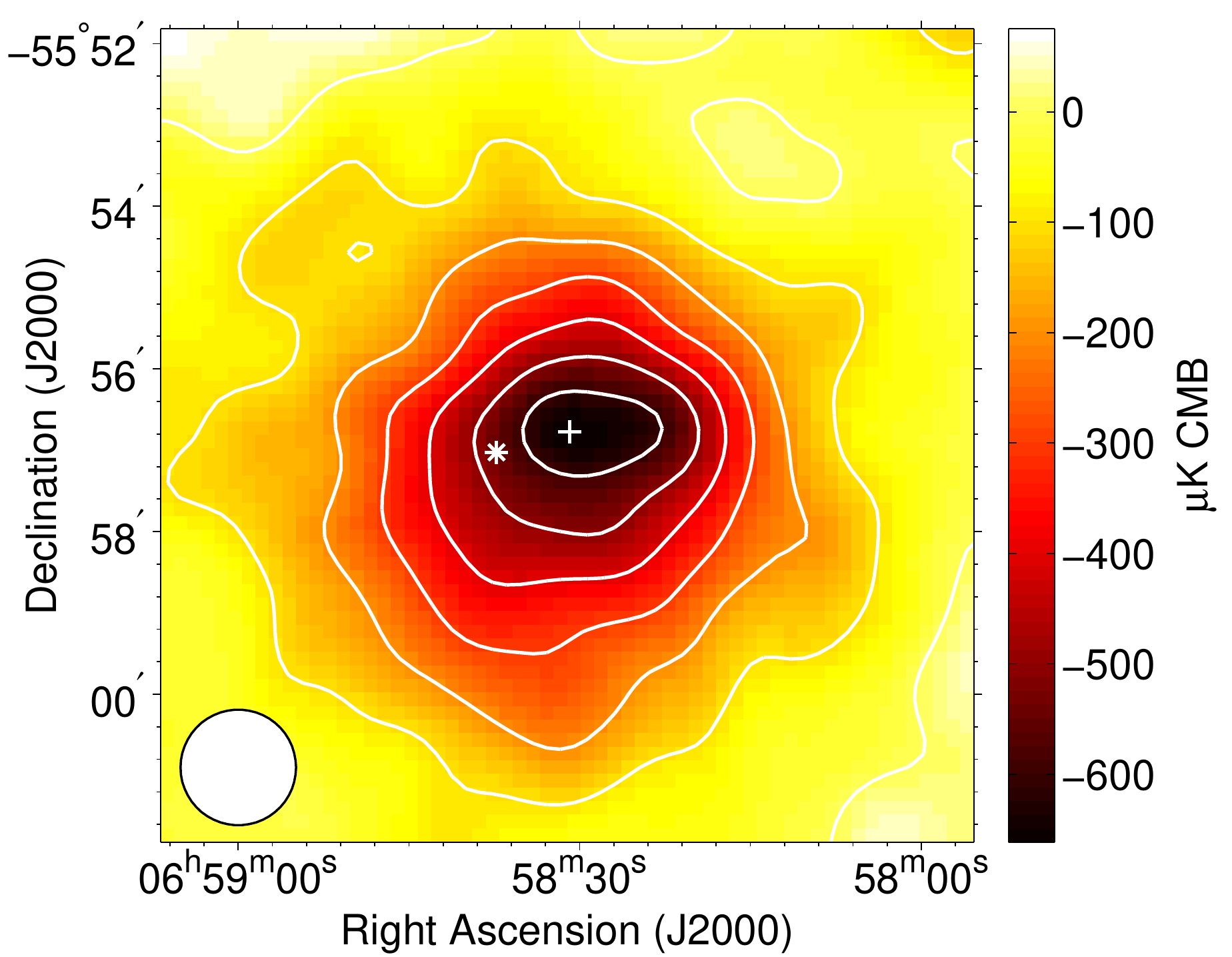}
  \caption[]{ Temperature map detail from
   Figure~\ref{FIG:bulletcoadd}, with color scale adjusted to the
   limits of the detail region, and a contour interval of 100~$\mu \rm{K}_{\rm{CMB}}$.
   The $+$ marker indicates the centroid
   position of the best-fit elliptical $\beta$ model, see
   \S\,\ref{subsec:fitting}. The $*$ marker indicates the position of
   the bright, dust obscured, lensed galaxy detected at 270~GHz by
   \markcite{wilson2008}{Wilson} {et~al.} (2008), see \S\,\ref{SEC:source_contributions}.}
  \label{FIG:bulletcoaddzoom}
\end{figure}

Figures \ref{FIG:bulletcoadd}~and~\ref{FIG:bulletcoaddzoom} show the
source-masked temperature map from our observations of the Bullet
cluster. The map has a resolution of 85\arcsec\ FWHM which results from the
combination of the 58\arcsec\ instrumental beam, the data reduction filters, and
a final 1\arcmin\ FWHM Gaussian smoothing of the map. The
source-masked map is shown in order to provide a more accurate representation of the
extended emission and cluster morphology.
The noise in the central region of the source-masked map
is 55~$\mu \rm{K}_{\rm{rms}}$ per 85\arcsec\ FWHM resolution
element.
Near the cluster center, the emission hints at elongation
in the East-West direction, which is along the axis between the main
and sub-cluster gas detected in the X-ray, see \S\,\ref{sec:xray}.
The more extended emission appears to be elongated
in the Northwest-Southeast direction, which is the major axis of the best-fit
elliptical $\beta$ model discussed in \S\,\ref{subsec:fitting}.
Figure~\ref{FIG:bulletcoaddzoom} shows the centroid position of the
best-fit elliptical $\beta$ model, and the position of the dust
obscured, lensed galaxy detected at 270~GHz by \markcite{wilson2008}{Wilson} {et~al.} (2008).
As discussed in \S\,\ref{SEC:source_contributions},
we see no evidence for emission from this source in our 150~GHz map.

Radial profiles for the unsmoothed source-masked and non-source-masked
maps are shown in Figure~\ref{FIG:radial_plots}. The source-masked map
has a signal-to-noise of 10 within the central 1\arcmin\ radius,
compared to 23 for the non-source-masked map, due to the fact that
source-masking allows more large-scale atmospheric fluctuation noise
to remain in the map.  However, the source-masked map preserves signal on
larger spatial scales than the non-source-masked map.  In both the
source-masked and non-source-masked maps, the sky intensity
distribution is filtered by the instrument beam and data reduction
pipeline described in \S\,\ref{SEC:datareduction}. We do not
renormalize the map amplitudes, since the source is extended and an
assumption would need to be made about the shape of the sky-brightness
distribution to do so. However, in order to accurately estimate
cluster parameters such as the central temperature decrement, a model
for cluster emission must be adopted, and the instrument beam and data
reduction filtering must be taken into account.

\subsection{Fit to Elliptical $\beta$ Model}\label{subsec:fitting}

We fit an elliptical $\beta$ model to the non-source-masked temperature map
to allow a straightforward comparison of cluster gas properties
derived from our measurements to those derived from X-ray
observations. In all analyses here we assume the cluster gas is
well-described by an isothermal $\beta$ model, and is in hydrostatic
equilibrium. These assumptions are unphysical in the case of the
Bullet cluster, which is a dynamically complex merging system where
the gas is separated from the rest of the mass~\markcite{clowe2006}({Clowe} {et~al.} 2006).
However, we find that with the sensitivity and spatial
resolution of the observations, these assumptions yield
an adequate description of the observed emission.

We model the three-dimensional radial profile of the
electron density with an isothermal $\beta$ model
\markcite{cavaliere1978}({Cavaliere} \& {Fusco-Femiano} 1978):

\begin{equation}
\label{EQN:betamodeldensity}
n_e(r) = n_{e0}\left( 1 + \frac{r^2}{r_c^2}\right)^{-3\beta/2}.
\end{equation}
Here, $n_{e0}$ is the central electron number density, $r_c$ is the
core radius of the gas distribution, and $\beta$ describes the
power-law index at large radii.

The radial surface temperature
profile of the SZE takes a simple analytic form:
\begin{equation}
\label{EQN:betamodeltemperature}
\Delta T_{\rm{SZ}} = \Delta
T_{0}\left( 1 + \frac{\theta^2}{\theta_c^2} \right)^{(1-3\beta)/2},
\end{equation}
where $\Delta T_{0}$ is the central temperature decrement, and
$\theta_{c} = r_c/D_A$ is the core radius divided by the angular-diameter distance. A similar form exists for the
X-ray surface brightness.

Because of the significant ellipticity in the measured SZE intensity
profile, we generalize the cluster gas model to be a spheroidal rather
than spherical function of the spatial coordinates:
\begin{equation}
\label{EQN:ellipticalbeta}
\Delta T_{\rm{SZ}} = \Delta T_{0}\left(1 + A + B \right)^{(1-3\beta)/2},
\end{equation}
with
$$
A = \frac{(\cos(\Phi)(X - X_{\rm{0}}) + \sin(\Phi)(Y - Y_{\rm{0}}))^2}{\theta_{\rm{c}}^2},
$$
$$
B = \frac{(-\sin(\Phi)(X - X_{\rm{0}}) + \cos(\Phi)(Y - Y_{\rm{0}}))^2}{(\eta \theta_{\rm{c}})^2}.
$$
Here $(X - X_{\rm{0}})$ and $(Y - Y_{\rm{0}})$ are angular offsets on the sky in the RA and
DEC directions, with respect to center positions $X_{\rm{0}}$ and
$Y_{\rm{0}}$. The axial ratio, $\eta$, is the ratio between the minor and major axis
core radii, $\Phi$ is the angle
between the major axis and the RA ($X$) direction.
$\Delta T_{0}$ is given by the gas pressure integrated along the
central line of sight through the cluster:
\begin{equation}
\label{EQN:integratedpressure} \frac{\Delta T_{0}}{T_{\rm{CMB}}} =
\frac{k_{\rm{B}} \sigma_T}{m_e c^2} \int f(x,T_e) n_e(l) T_e(l) dl
\end{equation}
where $x = h \nu / k T$, $f(x,T_e)$ describes the frequency dependence
of the SZE, $\sigma_T$ is the Thomson scattering
cross-section, and $T_{\rm{CMB}} = 2.728$ K.
For all results in this paper, we use the relativistic SZE spectrum $f(x,T_e)$
provided by ~\markcite{nozawa2000}{Nozawa} {et~al.} (2000), and neglect the kinematic effect.
At 150~GHz and $T_e = 13.9$~keV (see \S\,\ref{sec:xray}), this is a 9\% correction to
the non-relativistic value.

To accurately estimate $\beta$ model parameters for the cluster, the
instrument beam and data reduction filters, or transfer function, must
be applied to the model before comparing it with the data \markcite{benson2003,reese00}(see,
e.g., {Benson} {et~al.} 2003; {Reese} {et~al.} 2000). We characterize this transfer function
by creating a map from a simulated point source, convolved with the
instrument beam, and inserted into a noiseless timestream, similar to
the method described in \markcite{scott2008}{Scott} {et~al.} (2008). The point source transfer
function map $\mathcal{K}$ is then convolved with a simulated $\beta$
model cluster map $\mathcal{B}$ to generate a filtered model map
$\mathcal{B^\prime}$, which is a noiseless simulated APEX-SZ
observation of a $\beta$ model cluster. The filtered model map,
$\mathcal{B^\prime}$, is then differenced with the data map,
$\mathcal{M}$, and model parameters are estimated by minimizing a
$\chi^2$ statistic.

Simulating maps of many different cluster models is required for model
parameter fitting. Convolving the cluster model with the point source
transfer function map is much faster than processing each model
through the reduction pipeline. We find that the resulting simulated
maps from both methods agree sufficiently well to have negligible
effect on the parameter estimation results.

We use the unsmoothed non-source-masked map with $10\arcsec$
pixelization described in \S\,\ref{SEC:mapmaking} for all parameter
estimation described below, since this map has lower noise and a more
easily characterized transfer function than the source-masked map
shown in Figure~\ref{FIG:bulletcoadd}. Diffuse cluster emission is
more attenuated in the non-source-masked map, but this
is taken into account using the point source transfer function.

Map noise properties are assessed in the spatial frequency domain
using jackknife noise maps \markcite{sayers07,sayers08}(see Sayers 2007; {Sayers} {et~al.} 2009).
To estimate the noise covariance $C_n$, we assume that the noise is
stationary in the map basis. With this assumption, the Fourier
transform of the noise covariance matrix, $\widetilde{C}_n$, is
diagonal, and the diagonal elements are equal to the noise map power
spectral density (PSD). For each of 500 jackknife noise map
realizations, we find the two-dimensional Fourier transform, then
average over all realizations.
This averaged map PSD is the experimental estimate of the diagonal elements of
$\widetilde{C}_n$.
However, these jackknife maps do not include fluctuations due to the
primary CMB anisotropies.  We estimate the CMB signal covariance from
the WMAP5 best-fit power spectrum \markcite{nolta09}({Nolta} {et~al.} 2009) convolved with the point source
transfer function described earlier and add it to the jackknife noise
PSD to determine the total covariance matrix.

We construct a $\chi^2$ statistic for the
model fit using the transform of the filtered $\beta$ model,
$\widetilde{\mathcal{B^\prime}}$, and the transform of the
central $14\arcmin \times 14\arcmin$ portion of the data map
$\widetilde{\mathcal{M}}$ as:
\begin{equation}
\chi^2 = (\widetilde{\mathcal{M}}-\widetilde{\mathcal{B^\prime}})^T \widetilde{C}_n^{-1} (\widetilde{\mathcal{M}}-\widetilde{\mathcal{B^\prime}}).
\end{equation}
Using Markov Chain Monte Carlo (MCMC) methods, the likelihood,
$\mathcal{L} = e^{-\frac{1}{2} \chi^2}$, is sampled in the 7-dimensional
model parameter space and integrated to find
the marginal likelihood distributions of the $\beta$ model parameters.
The model parameter estimates and uncertainties that we report are the
maximum likelihood values and constant-likelihood 68\% confidence
intervals, respectively, of the marginal likelihood distributions.

The above approach to noise covariance estimation is chosen for its
simplicity and because we do not have enough linear combinations of
individual scan maps to fully sample the noise covariance matrix using
jackknife noise maps. But, the method relies on several simplifying
assumptions, including that the bolometer noise is stationary for each
100-s scan, the timestream noise is uncorrelated from scan to scan,
and the map coverage is uniform. Our map coverage is not actually
uniform, but we find through simulations of non-uniform Gaussian
noise maps that the $\chi^2$ statistic is not significantly affected.
In addition, the validity of the approach is tested by inserting
simulated clusters into the real timestream data; the simulated
cluster parameters are accurately recovered within the estimated
uncertainties.

Results of the $\beta$ model parameter estimation are given in
Table~\ref{TBL:betamodelfits}. Due to the degeneracy between the core
radius $\theta_{\rm{c}}$ and $\beta$ parameters, we assume a prior probability
density on $\beta$ of $1.04^{+0.16}_{-0.10}$, which is found from fits to ROSAT X-ray data
by \markcite{ota2004}{Ota} \& {Mitsuda} (2004). \markcite{hallman2007}{Hallman} {et~al.} (2007) find that in hydro/N-body simulations,
$\beta$ derived from fits to SZE profiles is higher than that from
X-ray, with $\beta_{\mathrm{SZ}}/\beta_{\mathrm{X}\mbox{-}\mathrm{ray}} = 1.21 \pm 0.13$ for fits
within $r_{500}$. We do not account for that factor here due to the
significant uncertainty in the X-ray derived $\beta$ value, but we
note that our SZE data prefer a higher value for $\beta$ than
the peak value of the prior. We further discuss this choice of prior in
\S\,\ref{subsec:mwte}.

The best-fit $\beta$ model fits the data well, with a reduced $\chi^2$
value of 1.008, and with 7219 degrees of freedom (DOF) has a probability
to exceed (PTE) of 31.5\%. The difference map between the data map,
$\mathcal{M}$, and the best-fit filtered $\beta$ model,
$\mathcal{B^\prime}$, shows no evidence of residual cluster structure
or point sources.

\begin{deluxetable}{llll}
\tablecaption{\label{TBL:betamodelfits} $\beta$ Model Fit Results}
\tablewidth{0pt}
\tablehead{
\colhead{Parameter} & \colhead{Description} & \colhead{Value} & \colhead{Uncertainty$^a$}
}
\startdata
$X_{\rm{0}}$ & RA centroid position &$06^{\rm h}58^{\rm m}30.86^{\rm s}$ (J2000) & $\pm 7.4\arcsec$ \\
$Y_{\rm{0}}$ & DEC centroid position & $-55\degree56\arcmin46.2\arcsec$ (J2000) & $\pm 7.3\arcsec$ \\
$\Delta T_{\rm{0}}$ & Central temperature decrement & $-771\ \mu$K$_{\rm{CMB}}$ & $\pm 71\ \mu$K$_{\rm{CMB}}$ \\
$y_0$ & Central Comptonization$^b$($T_e = 13.9$~keV) &  $3.31 \times 10^{-4}$ &  $\pm 0.30 \times 10^{-4}$ \\
$y_0$ & Central Comptonization$^b$($T_e = 10.6$~keV) &  $3.24 \times 10^{-4}$ &  $\pm 0.30 \times 10^{-4}$ \\
$\theta_{\rm{c}} $ & Core radius & $142 \arcsec$ & $\pm 18 \arcsec$ \\
$\eta$ & Ellipse minor/major core radius ratio & 0.889 & $\pm 0.072$ \\
$\Phi $ & Ellipse orientation angle & $-52\degree$ & $\pm 20\degree$ \\
$\beta$ & Power-law index & 1.15 & $\pm 0.13$ \\

\enddata
\tablenotetext{a}{Quoted uncertainties are 68\% confidence intervals
in the marginal likelihood distribution for each parameter.  The
uncertainty in $\Delta T_{\rm{0}}$ includes a statistical uncertainty
from the fit of $\pm 57\,\mu$K, and a $\pm 5.5\%$ flux calibration
uncertainty. The uncertainties in the centroid parameters $X_{\rm{0}}$
and $Y_{\rm{0}}$ include a $\pm 4\arcsec$ pointing uncertainty
and are given in units of arcseconds on the sky.}
\tablenotetext{b}{Central Comptonization, $y_0$, is a derived parameter,
assuming an electron temperature of 13.9 keV from \markcite{govoni2004}{Govoni} {et~al.} (2004)
and 10.6~keV from \markcite{zhang2006}{Zhang} {et~al.} (2006), an SZE observation frequency of 152~GHz, and
$T_{\rm CMB} = 2.728$~K. It is provided to facilitate comparisons with data at other
wavelengths.}

\end{deluxetable}

\begin{figure}[h!]\centering
\includegraphics[width=0.45\textwidth]{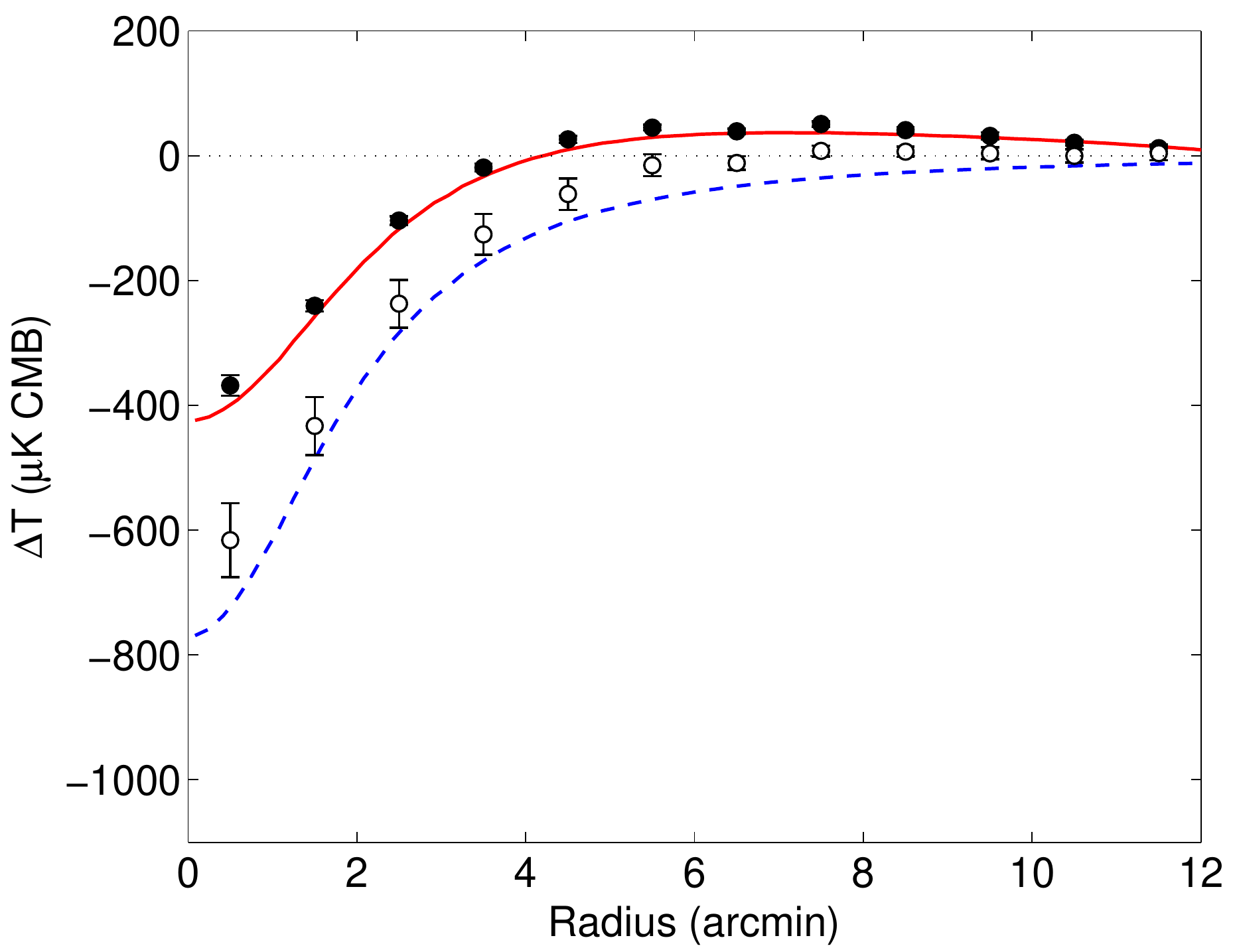}
  \caption[]{Radial profile of Sunyaev-Zel'dovich effect in the Bullet
  cluster compared to the best-fit $\beta$ model. Points with error
  bars are the SZE data binned in 1\arcmin\ radial bins, from the
  non-source-masked map, ($\mathcal{M}$, filled circles) and the
  source-masked map (open circles). The lines show the radial profile
  of the best-fit $\beta$ model, unfiltered ($\mathcal{B}$, blue
  dashed line) and after convolving with the instrument beam and
  non-source-masked data reduction filters ($\mathcal{B^\prime}$, red
  solid line). The non-source-masked map radial profile is reasonably
  well fit by the filtered $\beta$ model. The source-masked map
  preserves signal on larger spatial scales than the non-source-masked
  map, but still attenuates signal on scales exceeding the
  $4.75^\prime$ radius of the source masking. The source-masked data
  thus have a lower signal amplitude when compared with the unfiltered
  $\beta$ model, as expected.  See text for details.}
  \label{FIG:radial_plots}
\end{figure}

Radial profile plots of the best-fit $\beta$ model, $\mathcal{B}$, and
the filtered $\beta$ model map, $\mathcal{B^\prime}$, are shown in
Figure~\ref{FIG:radial_plots}.  Also plotted for comparison are the
radially binned data from the unsmoothed non-source-masked map
$\mathcal{M}$, used for model fitting, and the unsmoothed
source-masked map, used to visualize extended emission (without the
1\arcmin\ Gaussian smoothing used in
Figures~\ref{FIG:bulletcoadd}~\&~\ref{FIG:bulletcoaddzoom}).
Uncertainties in both sets of radially binned data are highly
correlated due to large-spatial-scale correlated noise in the
maps. The coincidence of the non-source-masked data ($\mathcal{M}$,
filled circles) and the filtered best-fit $\beta$ model
($\mathcal{B^\prime}$, red solid line) show that the data and best-fit
$\beta$ model are in good agreement. The source-masked map preserves
signal on larger spatial scales than the non-source-masked map, and is
useful for visualizing extended emission on larger spatial
scales. But, as expected, even the source-masked map attenuates signal
on scales exceeding the $4.75^\prime$ radius of the source masking, and thus has
a lower signal amplitude when compared with the unfiltered $\beta$ model
($\mathcal{B}$, blue dashed line).

\subsection{Radio and IR source contributions}
\label{SEC:source_contributions}

Radio sources associated with a galaxy cluster and background IR
galaxy sources can have a significant impact on the measurement of the
SZE emission at 150~GHz. We interpret the published results of observations
of the Bullet cluster at other frequencies and conclude that the measured
SZE decrement is not significantly contaminated.

The most important source of potential confusion is a bright, dust obscured,
lensed galaxy in the direction of the Bullet cluster recently reported
by \markcite{wilson2008}{Wilson} {et~al.} (2008). This source has a flux density of $13.5 \pm
0.5$~mJy at an observing frequency of 270~GHz, and is centered at
RA $06^{\rm h}58^{\rm m}37.31^{\rm s}$, DEC $-55\degree57\arcmin1.5\arcsec$ (J2000),
$\approx 56^{\prime \prime}$ to
the east of the measured SZE centroid position, see Figure~\ref{FIG:bulletcoaddzoom}. 
Assuming a spectral index
$\alpha=3$, where $S \propto \nu^\alpha$, we expect a flux density of
$1.94\,$mJy at $150\,$GHz corresponding to a temperature increment of
$\Delta T = 38\ \mu{\rm K_{CMB}}$ in the 1.5~arcmin$^2$ APEX-SZ beam solid angle.
This lensed
source is expected to be the dominant contribution to positive flux in
the direction of the cluster and we have repeated the $\beta$ model
fit taking it into account. We first add the source at its
measured position with the predicted 150~GHz flux to the SZE $\beta$
model and repeat the model fit. As expected, including the point
source results in a slightly ($\sim\sigma/3$) deeper decrement;
however, the $\chi^2$ of the model fit slightly increases.
We next allow the flux of the point source to vary along
with the other model parameters and find that values of positive flux
are a poorer fit than no source at all.
Therefore, we have no evidence for significant emission from this source
at 150~GHz. For the results in this paper, we use
cluster model parameters derived from fits that do not include this IR
source.

The Bullet cluster is also associated with a number of relatively
compact radio sources and one of the brightest cluster radio halos yet
discovered. However, these sources are predicted to produce negligible
temperature increments in the APEX-SZ beam when extrapolated to
150~GHz.

\markcite{liang2000}{Liang} {et~al.} (2000) report the detection of eight radio point sources
all of which have steeply falling spectra. Only two of these sources
were detected with ACTA at 8.8~GHz, and they were found to have flux
densities of $3.2 \pm 0.5$~mJy and $3.3 \pm 0.5$~mJy. The spectra of
these sources, measured between $4.9$ and $8.8\,$GHz are falling with
$\alpha = -0.93$ and $\alpha=-1.33$, respectively. Extrapolating to
$150\,$GHz, the flux of these sources are expected to be $0.24$ and
$0.07\,$mJy, corresponding to CMB temperature increments of $\Delta T
= 4.7$ and $1.4\,\mu{\rm K_{CMB}}$ in the 1.5~arcmin$^2$ APEX-SZ
beam solid angle.

The radio halo in the Bullet cluster is very luminous, but has a
characteristically steeply falling spectrum. \markcite{liang2000}{Liang} {et~al.} (2000)
measure the flux and spectra for the two main spatial components of
the halo. At $8.8\,$GHz, they find the two components to have fluxes
of $3.5$ and $0.55\,$mJy, with spectral indices of $\alpha=-1.3$ and
$\alpha=-1.4$, respectively. Extrapolating to $150\,$GHz, the combined flux from the
radio halo is expected to be $\sim 0.1\,$mJy. This emission is spread
over an area comparable to the size of the cluster and therefore
corresponds to a temperature increment $< 1\,\mu{\rm K_{CMB}}$ in the
APEX-SZ beam.

\subsection{Comparison with X-ray Data}\label{sec:xray}

X-ray emission in the ionized intracluster gas is dominated by thermal
bremsstrahlung. The X-ray surface brightness can be written
\begin{equation}
\label{xraysurfacebrightness}
 S_{X} = \frac{1}{4 \pi (1+z)^4} \int
n_{e} n_{i} \Lambda_{ei} dl,
\end{equation}
where $n_{e,i}$ are the electron and ion densities in this gas,
$\Lambda_{ei}$ is the X-ray cooling function, and the
integral is taken along the line of sight. The X-ray flux is
proportional to the line-of-sight integral of the square of the
electron density, resulting in emission that is more sensitive
to local density concentrations than the SZE emission.

\begin{figure}[h]\centering
\includegraphics[width=0.45\textwidth]{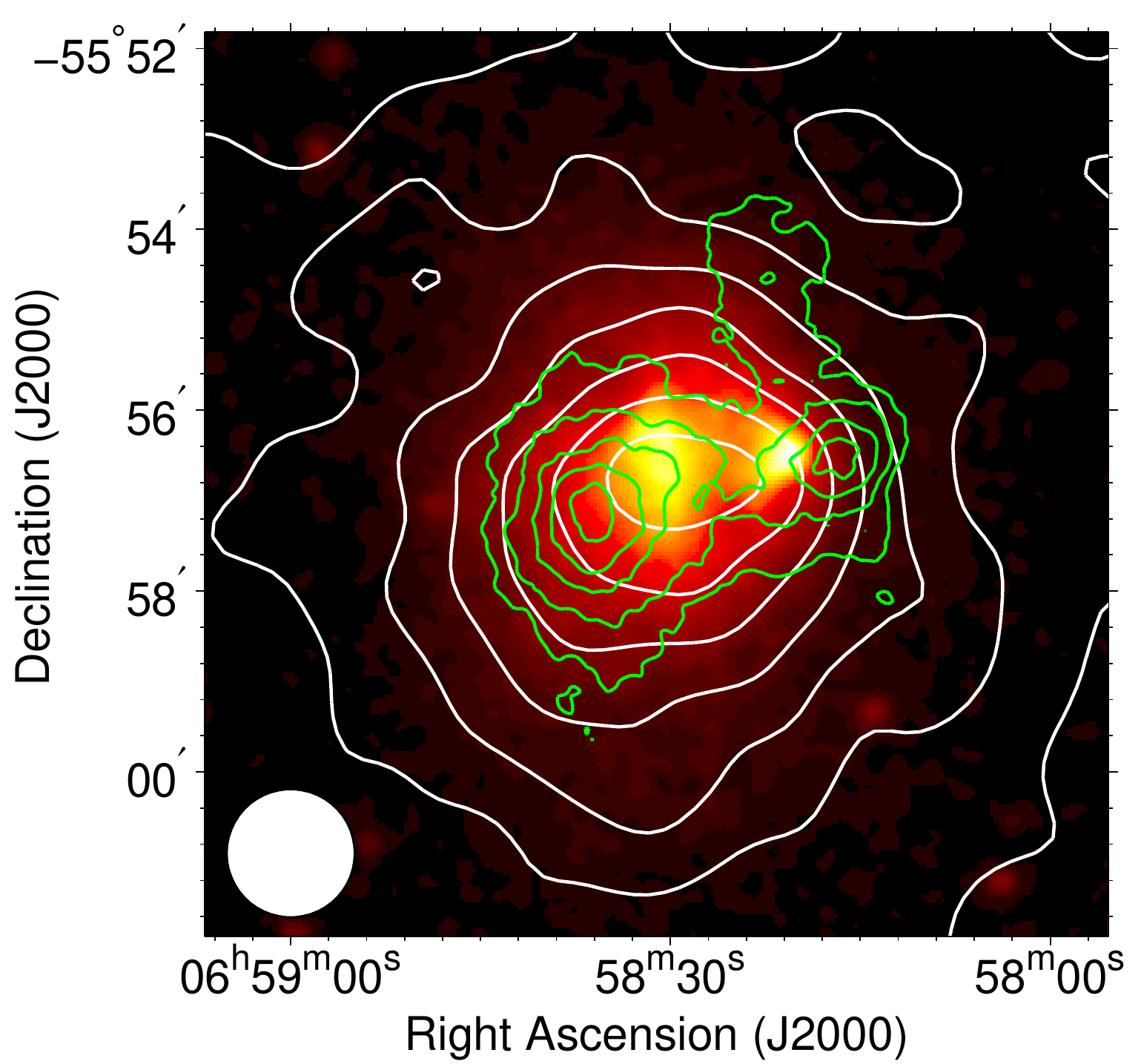}
  \caption[]{The SZE map of the Bullet system from this work, in white
  contours, overlaid on an X-ray map from XMM observations. The green
  contours show the weak lensing surface mass density reconstruction
  from \markcite{clowe2006}{Clowe} {et~al.} (2006). The SZE contour interval is 100~$\mu \rm{K}_{\rm{CMB}}$.}
  \label{FIG:bullet-overlay}
\end{figure}

The Bullet sub-cluster and the bow shock are apparent in the X-ray
surface brightness map shown in Figure~\ref{FIG:bullet-overlay}.
The SZE contour map of the Bullet cluster in
Figures~\ref{FIG:bulletcoadd}~\&~\ref{FIG:bulletcoaddzoom} is
overlaid on an X-ray map and weak lensing
surface mass density reconstruction from
\markcite{clowe2006}{Clowe} {et~al.} (2006).\footnote{Data are publicly available at
http://flamingos.astro.ufl.edu/1e0657/public.html.} The X-ray map is
made from XMM data (observation Id: 0112980201) extracted in the
[0.5-2]~keV band, corresponding to Bullet rest frame energies
where the X-ray cooling function for hot gas is relatively
insensitive to temperature. The map is smoothed with a 12\arcsec\
Gaussian kernel.

The SZE contours do not resolve the sub-cluster. However, an
elongation of the inner contours to the West suggests that a
contribution from it may be detected. The observed SZE map is
consistent with expectations, given the 85\arcsec\ resolution of
the SZE map, the different dependence of the X-ray and
SZE signals on gas density, and the mass and temperature difference between the two
merging components which predict a factor of $\sim10$ lower
integrated pressure from the sub-cluster. There is no evidence in the
SZE contours of a contribution from the lensed sub-mm bright galaxy
discussed in \S\,\ref{SEC:source_contributions}.

\subsection{Mass-Weighted Temperature}\label{subsec:mwte}

The combination of cluster SZE and X-ray measurements can be used to place
constraints on the thermal structure of the intracluster gas.
The SZE intensity is proportional to the product of the electron density and the
electron temperature along the line of sight, see equation~(\ref{EQN:integratedpressure}). Therefore, if
the electron density is known from another measurement, the SZE
can be used to measure a mass-weighted temperature.
For simplicity, we assume here that the intracluster gas is isothermal, but a more detailed
comparison of the SZE surface brightness and projected density could be used to
constrain the thermal structure in the cluster.

We perform this calculation with two different descriptions of the intracluster gas density.
First, we model the spatial distribution of the intracluster gas as a
spherical $\beta$ model following equation~(\ref{EQN:betamodeldensity}).
We use $\beta$ model parameters from \markcite{ota2004}{Ota} \& {Mitsuda} (2004), derived from ROSAT HRI
($\sim 2\arcsec$ resolution) measurements of the inner $6\arcmin$ radius of the
Bullet cluster: $\beta = 1.04^{+0.16}_{-0.10}$, $\theta_c =
112.5^{+15.6}_{-10.4} \arcsec$, and $n_{e0} = 7.2^{+0.3}_{-0.3}
\times 10^{-3}$ cm$^{-3}$.

We construct an X-ray derived SZE surface brightness model from
the $\beta$ model electron surface density profile using
equation~(\ref{EQN:integratedpressure}). To account for $\beta$ model
uncertainties, we incorporate the values and
uncertainties for $\beta$, $\theta_c$, and $n_{e0}$ as independent Gaussian
priors. We then use the analysis method described in
\S\,\ref{subsec:fitting} to minimize $\chi^2$ on the difference
between the X-ray derived SZE model, convolved with the point source
transfer function, and the APEX-SZ non-source-masked data. The free
parameters in the fit are the three $\beta$ model parameters, the mass-weighted electron
temperature $T_{mg}$, and the relative map alignment in RA and DEC.

We find $T_{mg}=11.4 \pm 1.4\,$keV after marginalizing over the other
parameters in the fit and including the SZE
flux calibration uncertainty and the
effect of the APEX-SZ band center frequency uncertainty on the
relativistic SZE spectrum $f(x,T_{mg})$. The reduced $\chi^2$ of the
best fit model is $1.008$ with an associated PTE of $31.3\%$,
indicating that the spherical $\beta$ density model and the assumption
of isothermality produce an acceptable fit to the data.

Given the complex morphology of this merging system, the validity of
the spherical $\beta$ model is questionable. We therefore repeat the
determination of the mass-weighted temperature by directly comparing
X-ray measurements of the projected intracluster gas density with the
measured SZE signal in order to produce a less model-dependent
measurement of the mass-weighted temperature.
We make use of the publicly
available\footnote{http://flamingos.astro.ufl.edu/1e0657/public.html}
electron surface density map derived from Chandra X-ray satellite data presented
in \markcite{clowe2006}{Clowe} {et~al.} (2006).
Using the same analysis as above, and marginalizing over the relative
map alignment parameters, we find mass-weighted electron temperature
$T_{mg} = 10.8 \pm 0.9\,$keV. The fit to the data is again good, with a
reduced $\chi^2= 1.037$ and a PTE of 20.6\%. This is in excellent
agreement with the value $T_{mg}=11.4 \pm 1.4 \,$keV found from the above
$\beta$ model analysis.

Given the complex dynamics in the Bullet cluster, there have been several
studies of the temperature structure
\markcite{finoguenov2005, markevitch2006,andersson2007}(e.g., {Finoguenov}, {B{\"o}hringer}, \&  {Zhang} 2005; {Markevitch} 2006; {Andersson}, {Peterson}, \&  {Madejski} 2007).
There have also been several published results for the spectroscopic
temperature within annuli about the cluster center.
Chandra data was used by \markcite{govoni2004}{Govoni} {et~al.} (2004) to determine a spectroscopic
X-ray temperature of $T_{spec}=13.9 \pm 0.7\,$keV within $0.75\,$Mpc of the
cluster center.
From the analysis of XMM data within an annulus of $0.14-0.7\,$Mpc
radius, \markcite{zhang2006}{Zhang} {et~al.} (2006) find a temperature of $T_{spec}=10.6\pm 0.2\,$keV.
Analyzing the combination of XMM and RXTE data, \markcite{petrosian06}{Petrosian}, {Madejski}, \&  {Luli} (2006), find
$T_{spec}=12.1\pm0.2\,$keV within a radius of $0.95\,$Mpc.
The published X-ray spectroscopic temperatures span a range much larger
than the stated uncertainties in the measurements.
Given the complex thermal structure for the cluster, and the presence of
gas at temperatures corresponding to energies at or above the upper limits
of the Chandra and XMM energy response, the variation in the measured
X-ray spectroscopic temperature is not surprising.
The mass-weighted temperature found with APEX-SZ falls near the lowest of the
reported X-ray spectroscopic temperatures.
However, we do not expect exact agreement between the mass-weighted and
spectroscopic temperatures.
Using Chandra data for a sample of 13 relaxed clusters, \markcite{vikhlinin2006}{Vikhlinin} {et~al.} (2006)
find that, due to the presence of thermal structure in the intracluster gas, the
X-ray spectroscopic temperature is typically a factor of
$T_{spec}/T_{mg}=1.11 \pm 0.06$
larger than the X-ray derived mass-weighted electron temperature.
This is consistent with the simulation results of \markcite{nagai07}{Nagai}, {Vikhlinin}, \& {Kravtsov} (2007) who find
$T_{spec}/T_{mg} \approx 1.14$ for relaxed clusters and
$T_{spec}/T_{mg} \approx 1.12$, with a somewhat larger scatter,
for unrelaxed systems.
Naively applying this correction to the published X-ray spectroscopic
temperatures, we infer results for mass-weighted temperatures that bracket the
APEX-SZ result.

\subsection{Gas Mass Fraction}\label{sec:gmass}

Using the SZE measurements, we construct a model for the
intracluster gas distribution which, when combined with X-ray
measurements, can be used to determine the gas mass, total mass, and
therefore gas mass fraction of the cluster. The gas mass is estimated by
integrating a spheroidal model for the cluster gas, following
\markcite{laroque2006}{LaRoque} {et~al.} (2006).

Several assumptions about the model must be made to estimate the gas
mass. We assume that the cluster gas is isothermal in order to convert
pressure to density. We also assume spheroidal symmetry for the gas
distribution in order to convert the two-dimensional SZE integrated
pressure measurement to a three-dimensional gas distribution.

We consider two simple cases, an oblate spheroid generated by rotation
about the minor axis and a prolate spheroid generated by rotation
about the major axis, where the symmetry axis is in the sky plane.
The gas mass, under these assumptions, becomes:
\begin{eqnarray}
\label{EQN:gasmass}
\lefteqn{M_{\rm{gas}}(r) =
8 \mu_e n_{e0} m_p {D_{A}}^3 \int_0^{r/D_{A}} dX \, dY \, dZ } \nonumber \\
& & \left(1+\left(\frac{X}{\theta_{\rm{c}}}\right)^2+ \right.
\left.\left(\frac{Y}{\eta\theta_{\rm{c}}}\right)^2 +
\left(\frac{Z}{\zeta\theta_{\rm{c}}}\right)^2 \right)^{-3\beta/2}
\end{eqnarray}
where $\mu_e$ is the nucleon/electron ratio, taken to be
1.16~\markcite{grego2001}({Grego} {et~al.} 2001). The factor of eight is due to integrating over
only one octant of the spheroid. The factor $\zeta$ is set to unity in
the case of oblate spheroidal symmetry, while in the case of prolate spheroidal
symmetry, $\zeta$ is set to $\eta$.

The total cluster mass is estimated by assuming hydrostatic
equilibrium  and integrating the inferred gas distribution
\markcite{grego2000}({Grego} {et~al.} 2000) to find:
\begin{equation}
\label{EQN:hydroeq}
 \rho_{\rm{total}} = -\frac{kT_{\rm{e}}}{4\pi G \mu m_{\rm{p}}} \nabla^2 \ln \rho_{\rm{gas}}.
\end{equation}
Here $\mu$ is the mean molecular weight of the intracluster gas,
which is assumed to be 0.62 \markcite{zhang2006}({Zhang} {et~al.} 2006). Using
equations~(\ref{EQN:gasmass}) \&~(\ref{EQN:hydroeq}), and our model parameters in
Table~\ref{TBL:betamodelfits}, we calculate the gas mass, total
mass, and gas mass fraction for the cluster.
In Table~\ref{TBL:massestimates}, we give these results.
The gas mass fraction results for a prolate gas
distribution model are $\sim3\%$ larger than those for
an oblate model, while the total mass and gas mass are $\lesssim18\%$ larger.
We quote only the oblate spheroidal results.

We calculate our results within two different radii. The first is
the radius of the cluster at which its mean density is equal to 2500
times the critical density at the redshift of the cluster,
$r_{2500}$. The second radius is 1.42~Mpc, which is the same radius
used by \markcite{zhang2006}{Zhang} {et~al.} (2006) for their gas mass fraction calculation.
This will allow for a more direct comparison to their result, and is
also near where our measured SZE radial profile has unity signal to
noise.
For all results, we assume a $\Lambda$CDM cosmology,
with $h = 0.7$, $\Omega_{\rm{m}} = 0.27$, and $\Omega_{\Lambda} =
0.73$.
The results of this analysis are summarized in Table~\ref{TBL:massestimates}.  
Under the assumption of a $10.6\,$keV mass-weighted temperature (the lowest of the published
X-ray spectroscopic temperatures and near our mass-weighted temperature results 
in \S\,\ref{subsec:mwte}), we find gas mass fractions 
$f_g=0.216 \pm 0.031$ and $0.179 \pm 0.036$ within $r_{2500}$ and 1.42~Mpc, respectively.
The fact that the computed gas fraction in the central region significantly 
exceeds the cosmic average determined by WMAP5 \markcite{dunkley09}($f_g=0.165\pm 0.009$, {Dunkley} {et~al.} 2009), 
and a lower value observed in relaxed clusters 
\markcite{vikhlinin09}($f_g \simeq 0.12$, see, e.g., {Vikhlinin} {et~al.} 2009), is likely due to deviations of the 
intracluster gas from isothermal hydrostatic equilibrium.  
On larger scales, baryon fractions produced for the range of reported 
X-ray temperatures
bracket the published results using X-ray and weak lensing data.
\markcite{bradac2006}{Brada{\v c}} {et~al.} (2006) measure a gas
mass fraction $f_g=0.14 \pm 0.03$ by comparing the gas mass
calculated from Chandra X-ray measurements to weak lensing total
mass measurements in a $4.9\arcmin \times 3.2\arcmin$ box roughly
centered around the cluster. \markcite{zhang2006}{Zhang} {et~al.} (2006), measured a gas mass
fraction $f_g=0.161 \pm 0.018$ within a radius of 1.42~Mpc.
Despite the limitations of applying a hydrostatic equilibrium model 
to this merging cluster, the APEX-SZ results for the
gas mass fraction are in good agreement with previous work.

\begin{deluxetable}{ccccccc}
\tabletypesize{\tiny}
\tablecaption{\label{TBL:massestimates} Mass Estimates for the Bullet Cluster System}
\tablewidth{0pt}
\tablehead{
\colhead{$T_e (keV)^a$} & \colhead{Mean Overdensity} & \colhead{$r_{\rm{int}}$ (\arcmin)$^b$} & \colhead{$r_{\rm{int}}$ (Mpc)$^c$} & \colhead{Gas Mass Fraction} & \colhead{Gas Mass ($10^{14} M_{\sun}$)} & \colhead{Total Mass ($10^{14} M_{\sun}$)}
}
\startdata
$13.9 \pm 0.7$ & $2506 \pm 233$ & 2.77 & 0.739 & $0.124 \pm 0.022$ & $0.944 \pm 0.105$ & $7.56 \pm 0.70$ \\
$13.9 \pm 0.7$ & $961 \pm 98$ & 5.32 & 1.42  & $0.106 \pm 0.024$ & $2.20 \pm 0.33$ & $20.6 \pm 2.1 $ \\
$10.6 \pm 0.2$ & $2521 \pm 230$ & 2.15 & 0.572 & $0.216 \pm 0.031$ & $0.765 \pm 0.072$ & $3.54 \pm 0.32$ \\
$10.6 \pm 0.2$ & $734 \pm 66$ & 5.32 & 1.42  & $0.179 \pm 0.036$ & $2.83 \pm 0.40$ & $15.7 \pm 1.4 $ \\
\enddata
\tablecomments{Two different isothermal electron temperatures are assumed, in order to bracket the range of X-ray spectroscopic temperatures
reported in the literature. The top two rows assume an isothermal electron temperature of $13.9 \pm 0.7$ keV.  The bottom two rows assume an isothermal electron temperature of $10.6 \pm 0.2$ keV.  For each electron temperature, we integrate to $r_{2500}$, the radius within which the mean cluster density is 2500
times greater than the critical density at the redshift of the cluster.
For each electron temperature, we also integrate to a fixed radius of 1.42~Mpc, allowing a direct comparison
to results in \markcite{zhang2006}{Zhang} {et~al.} (2006). This radius is also near where our
measured SZE radial profile has unity signal to
noise. For all results, we assume a $\Lambda$CDM cosmology,
with $h = 0.7$, $\Omega_{\rm{m}} = 0.27$, and $\Omega_{\Lambda} =
0.73$. Uncertainties in the gas mass and gas mass fraction include
a $\pm 5.5\%$ SZE flux calibration uncertainty.}
\tablenotetext{a}{Isothermal electron temperature}
\tablenotetext{b}{Angular integration radius.}
\tablenotetext{c}{Physical integration radius.}
\end{deluxetable}


\section{Conclusions}
\label{SEC:conclusions} Measurements of the SZE provide a robust and
independent probe of the intracluster gas properties in galaxy
clusters. The APEX-SZ $150\,$GHz observations detect the Bullet system
with $23\,\sigma$ significance within the central 1\arcmin\ radius of the
SZE centroid position.
We do not expect to see a resolved signal from the Bullet
sub-cluster in the $150\,$GHz 85\arcsec\ FWHM resolution SZE maps,
and no obvious feature, such as a secondary peak, is present.
We expect no significant contamination of the observed SZE decrement
due to radio sources, and there is no evidence for
significant contamination by a known bright lensed dusty galaxy.

We process an elliptical $\beta$ model through the observation
transfer function and fit it to the measured temperature decrement map.
We also measure the cluster mass-weighted
electron temperature and gas mass fraction with the SZE data.
Combining the APEX-SZ map with a map of projected electron surface density from
Chandra X-ray observations, we determine the mass-weighted temperature of the
cluster gas to be $T_{mg}=10.8 \pm 0.9\,$keV.
This value is consistent with the lowest X-ray spectroscopic temperatures reported
for this cluster and should be less sensitive to the details of the cluster
thermal structure.
The derived baryon fraction is also found to be in reasonable agreement with
previous X-ray and weak lensing determinations.

Throughout this work, we make an assumption of isothermal cluster gas.
Clearly, incorporating thermal structure, measured by X-ray
observations, in the analysis of the SZE data would improve the
determination of the gas distribution and gas mass fraction.
Ultimately, a more sophisticated analysis could be implemented that
combines X-ray, SZE, and weak lensing data and relaxes assumptions of
hydrostatic equilibrium between the gas and dark matter components of
the cluster. This is particularly important for a detailed
understanding of actively merging systems such as the Bullet cluster.

\acknowledgments

We thank the staff at the APEX telescope site, led by David
Rabanus and previously by Lars-\AA ke Nyman, for their dedicated and exceptional support. We also thank
LBNL engineers John Joseph and Chinh Vu for their work on the
readout electronics. APEX-SZ is funded by the National Science
Foundation under Grant Nos.\ AST-0138348 \& AST-0709497.
Work at LBNL is supported
by the Director, Office of Science, Office of High Energy and
Nuclear Physics, of the U.S. Department of Energy under Contract No.
DE-AC02-05CH11231. Work at McGill is supported by the Natural Sciences
and Engineering Research Council of Canada and the Canadian Institute
for Advanced Research. RK acknowledges partial financial support from
MPG Berkeley-Munich fund. NWH acknowledges support from an Alfred P. Sloan
Research Fellowship.

\bibliography{}

\end{document}